\documentclass[12pt]{article}
\usepackage{graphicx}
\usepackage{lscape}
\usepackage{citesort}
\usepackage{amssymb}
\usepackage{appendix}
\usepackage{multirow}

\begin{document}

\def\Tr{\mbox{Tr}}
\def\figt#1#2#3{
        \begin{figure}
        $\left. \right.$
        \vspace*{-2cm}
        \begin{center}
        \includegraphics[width=10cm]{#1}
        \end{center}
        \vspace*{-0.2cm}
        \caption{#3}
        \label{#2}
        \end{figure}
	}
	
\def\figb#1#2#3{
        \begin{figure}
        $\left. \right.$
        \vspace*{-1cm}
        \begin{center}
        \includegraphics[width=10cm]{#1}
        \end{center}
        \vspace*{-0.2cm}
        \caption{#3}
        \label{#2}
        \end{figure}
                }

\def\ds{\displaystyle}
\def\beq{\begin{equation}}
\def\eeq{\end{equation}}
\def\bea{\begin{eqnarray}}
\def\eea{\end{eqnarray}}
\def\beeq{\begin{eqnarray}}
\def\eeeq{\end{eqnarray}}
\def\ve{\vert}
\def\vel{\left|}
\def\ver{\right|}
\def\nnb{\nonumber}
\def\ga{\left(}
\def\dr{\right)}
\def\aga{\left\{}
\def\adr{\right\}}
\def\lla{\left<}
\def\rra{\right>}
\def\rar{\rightarrow}
\def\lrar{\leftrightarrow}  
\def\nnb{\nonumber}
\def\la{\langle}
\def\ra{\rangle}
\def\ba{\begin{array}}
\def\ea{\end{array}}
\def\tr{\mbox{Tr}}
\def\ssp{{\Sigma^{*+}}}
\def\sso{{\Sigma^{*0}}}
\def\ssm{{\Sigma^{*-}}}
\def\xis0{{\Xi^{*0}}}
\def\xism{{\Xi^{*-}}}
\def\qs{\la \bar s s \ra}
\def\qu{\la \bar u u \ra}
\def\qd{\la \bar d d \ra}
\def\qq{\la \bar q q \ra}
\def\quu{\la \bar{q}_1 q_1 \ra}
\def\qdd{\la \bar{q}_2 q_2 \ra}
\def\qss{\la \bar{q}_3 q_3 \ra}
\def\GG{\langle g_s^2 G^2 \rangle}
\def\q{\gamma_5 \not\!q}
\def\x{\gamma_5 \not\!x}
\def\g5{\gamma_5}
\def\sb{S_Q^{cf}}
\def\sd{S_d^{be}}
\def\su{S_u^{ad}}
\def\sbp{{S}_Q^{'cf}}
\def\sdp{{S}_d^{'be}}
\def\sup{{S}_u^{'ad}}
\def\ssp{{S}_s^{'??}}

\def\sig{\sigma_{\mu \nu} \gamma_5 p^\mu q^\nu}
\def\fo{f_0(\frac{s_0}{M^2})}
\def\ffi{f_1(\frac{s_0}{M^2})}
\def\fii{f_2(\frac{s_0}{M^2})}
\def\O{{\cal O}}
\def\sl{{\Sigma^0 \Lambda}}
\def\es{\!\!\! &=& \!\!\!}
\def\ap{\!\!\! &\approx& \!\!\!}
\def\ar{&+& \!\!\!}
\def\ek{&-& \!\!\!}
\def\kek{\!\!\!&-& \!\!\!}
\def\cp{&\times& \!\!\!}
\def\se{\!\!\! &\simeq& \!\!\!}
\def\eqv{&\equiv& \!\!\!}
\def\kpm{&\pm& \!\!\!}
\def\kmp{&\mp& \!\!\!}
\def\mcdot{\!\cdot\!}
\def\erar{&\rightarrow&}


\def\simlt{\stackrel{<}{{}_\sim}}
\def\simgt{\stackrel{>}{{}_\sim}}


\title{
         {\Large
                 {\bf
$\gamma^\ast N \to N^\ast (1520)$ transition form factors in light cone QCD
sum rules
                 }
         }
      }

\author{\vspace{1cm}\\
{\small T. M. Aliev \thanks {
taliev@metu.edu.tr}~\footnote{Permanent address: Institute of
Physics, Baku, Azerbaijan.}\,\,,
M. Savc{\i} \thanks
{savci@metu.edu.tr}} \\
{\small Physics Department, Middle East Technical University,
06531 Ankara, Turkey }}

\date{}

\begin{titlepage}
\maketitle
\thispagestyle{empty}

\begin{abstract}

The $\gamma^\ast N \to N^\ast (1520)$ reaction is studied in framework of
the light cone QCD sum rules method. The sum rules for the multipole form
factors, namely, for the magnetic dipole $G_M(Q^2)$, electric quadrupole
$G_E(Q^2)$, and Coulomb quadrupole $G_C(Q^2)$,  form factors responsible for
for the $\gamma^\ast N \to N^\ast (1520)$ transition are derived. Using
these sum rules, the helicity amplitudes are presented as the combinations of
these multipole form factors, and their $Q^2$ dependences are studied.
We compare our results on the helicity amplitudes with those of the
spectator model predictions, and with the existing experimental data.

\end{abstract}

~~~PACS numbers: 11.55.Hx, 13.30.Ce, 13.40.Gp, 14.20.Jn

\end{titlepage}

\section{Introduction}

The study of the electromagnetic structure of the baryons proves to be an
important source for understanding their inner structure, as well as for
obtaining information on strong interaction in the confinement region. The
electromagnetic structure of baryons is described by the form factors whose
experimental and theoretical investigations have been continuing for a long
time up today. Operation of new electron beam facilities at
Jefferson Laboratory (JLAB), Mainz Microton (MAMI) at Mainz, and
MIT/Bates open new horizons in study of the form factors of nucleons and
their excitations. The nucleon excitations are studied in reactions such as
scattering of electron beams or target photons from nucleons, i.e.,
$e(\gamma^\ast) N \to e(\gamma^\ast) N^\ast$, where $N^\ast$ is the nucleon
excitation and $\gamma^\ast$ is the virtual photon. Obviously, more
information can be obtained about the structure of baryons at small
distances with the increasing photon virtuality.

It is hoped that the new electron beam facilities would definitely allow
collecting large amount of more precise and complete data in the studies of the
electro-excitations of nucleon resonances. Along these lines essential
information has already been obtained, in particular, for the spin-3/2 and
negative parity $N^\ast(1520)$ state in the range up to $Q^2=4.5~GeV^2$.
Progress in the experiments has simulated new significant developments in
understanding the dynamics of QCD at large distance from theoretical side
(for more detail see \cite{Rbozo01}).

In the present work we study the $\gamma^\ast N \to N^\ast(1520)$
transition form factors in framework of the light cone QCD sum rules (LCSR)
\cite{Rbozo02}. This transition was studied within the framework of the
nonrelativistic and relativistic quark model (see \cite{Rbozo03,Rbozo04} and
references therein), single quark transition model \cite{Rbozo05}, and the
covariant spectator quark model \cite{Rbozo06}, respectively.

The experimental results for the transition form factors were obtained from
CLAS data \cite{Rbozo07,Rbozo08}. Recently new data has been obtained in a
wider $Q^2$ domain, which surely has higher accuracy compared to the
previous experimental results.

The paper is organized as follows: In section 2 the sum  rules for the form
factors responsible for the $\gamma^\ast N \to N^\ast(1520)$ transition are
obtained within the LCSR method. In this section we also
present the expressions for the helicity amplitudes which describe this
transition. In section 3 numerical results for the transition form factors,
as well as for the helicity amplitudes are presented. This section contains
also discussions and conclusions. 

\section{Form factors of $\gamma^\ast N \to N^\ast(1520)$
transition in LCSR}

In order to derive the light cone QCD sum rules for the $\gamma^\ast N \to
N^\ast(1520)$ we introduce the following correlation function:
\bea
\label{ebozo01}
\Pi_{\alpha\mu}(p,q) = i \int d^4x e^{iqx} \lla 0 \vel T \Big\{\eta_\alpha(0)
j_\mu^{el}(x) \Big\} \ver N(p) \rra~,
\eea
where $\eta_\alpha$ is the interpolating current of the spin-3/2 positive
parity baryons, and
$j_\mu^{el}$ is the electromagnetic current defines as $j_\mu^{el} = e_u
\bar{u} \gamma_\mu u + e_d \bar{d} \gamma_\mu d$. The 
interpolating current of the spin-3/2 baryons is given by the following
expression:
\bea
\label{ebozo02}
\eta_\alpha = \varepsilon^{abc} \Big\{ \left( q_1^{aT} C \sigma_{\rho\lambda}
q_2^b \right) \sigma^{\rho\lambda} \gamma_\mu q_3^c -
\left( q_1^{aT} C \sigma_{\rho\lambda}           
q_3^b \right) \sigma^{\rho\lambda} \gamma_\mu q_2^c \Big\}~,
\eea
where $q_1=q_3=u$, $q_2=d$ for the excited proton, and
$q_1=q_3=d$, $q_2=u$ for the excited neutron states, respectively;
and $a$, $b$, and $c$ are the color indices and $C$ is the charge
conjugation operator.

We start our discussion by considering the hadronic transition involving
$N^\ast(1520)$. Within the framework of LCSR method this is implemented by
calculating the correlation function from the hadronic and quark-gluon
sides, respectively, and then equating these representations. In calculating
the correlation function from the hadronic side, the total set of hadrons
which carry the same quantum numbers of the interpolating current should be
inserted in it. The contributions of the lowest spin-3/2 negative and
positive parity states can be obtained by saturating the hadronic part of
the correlation function with them, from which we get:
\bea
\label{ebozo03}
\Pi_{\alpha\mu} \es {1\over m_{\mbox{\tiny${3\over 2}^-$}} -
p^{\prime 2}} \la 0 \vel \eta_\alpha \ver
\mbox{\small ${3\over 2}^-$} (p^{\prime}) \ra \la \mbox{\small ${3\over
2}^-$} (p^{\prime}) \vel
j_\mu^{el} \ver N(p) \ra~, \nnb \\
\ar {1\over m_{\mbox{\tiny${3\over 2}^+$}} -
p^{\prime 2}} \la 0 \vel \eta_\alpha \ver  
\mbox{\small ${3\over 2}^+$} (p^{\prime}) \ra \la \mbox{\small ${3\over
2}^+$} (p^{\prime}) \vel             
j_\mu^{el} \ver N(p) \ra~, \nnb \\
\ar {1\over m_{\mbox{\tiny${1\over 2}^-$}} -
p^{\prime 2}} \la 0 \vel \eta_\alpha \ver  
\mbox{\small ${1\over 2}^-$} (p^{\prime}) \ra \la \mbox{\small ${1\over
2}^-$} (p^{\prime}) \vel             
j_\mu^{el} \ver N(p) \ra~, \nnb \\
\ar {1\over m_{\mbox{\tiny${1\over 2}^+$}} -
p^{\prime 2}} \la 0 \vel \eta_\alpha \ver  
\mbox{\small ${1\over 2}^+$} (p^{\prime}) \ra \la \mbox{\small ${1\over
2}^+$} (p^{\prime}) \vel
j_\mu^{el} \ver N(p) \ra~. 
\eea
It should be noted here that the current $\eta_\alpha$ interacts not only
with spin-3/2 states, but also spin-1/2 states, and therefore we consider
their contributions on Eq. (\ref{ebozo03}).

We start by considering the last two matrix elements which correspond to the
contributions of the spin--1/2 states. In the general case, the matrix element
$\la 0 \vel \eta_\alpha \ver \mbox{\small ${1\over 2}$}^+ (p^{\prime}) \ra$ 
is determined as follows:
\bea
\label{ebozo04}
\la 0 \vel \eta_\alpha \ver \mbox{\small ${1\over 2}$}^+ (p^{\prime}) \ra
= (A p_\alpha^\prime + B \gamma_\alpha ) u(p^\prime)~.
\eea
Multiplying both sides of Eq. (\ref{ebozo04}) and using the fact that
$\gamma_\alpha \eta^\alpha=0$, we immediately obtain that:
\bea
\label{ebozo05}
\la 0 \vel \eta_\alpha \ver \mbox{\small ${1\over 2}$}^+ (p^{\prime}) \ra =
B \Big( \gamma_\alpha - {4\over m_{\mbox{\tiny${1\over 2}^+$}}}
p_\alpha^\prime \Big)u(p^\prime)~.
\eea
Similarly, from the parity consideration the matrix element
$\la 0 \vel \eta_\alpha \ver \mbox{\small ${1\over 2}$}^- (p^{\prime}) \ra$
is determined as:
\bea
\label{ebozo06}
\la 0 \vel \eta_\alpha \ver \mbox{\small ${1\over 2}$}^- (p^{\prime}) \ra = 
B \gamma_5 \Bigg( \gamma_\alpha - {4\over m_{\mbox{\tiny${1\over 2}^-$}}}  
p_\alpha^\prime \Bigg) u(p^\prime)~.
\eea
It follows from Eqs. (\ref{ebozo05}) and (\ref{ebozo06}) that the
contributions of the spin-1/2 states are proportional to either
$p_\alpha^\prime$ or $\gamma_\alpha$, which we will use later.

The remaining matrix elements are determined as follows:
\bea
\label{ebozo07}
\la \mbox{\small ${1\over 2}^+$} (p^{\prime}) \vel j_\mu^{el} \ver
\mbox{\small ${1\over 2}^+$} (p) \ra =
\bar{u}(p^\prime) \Bigg[\gamma_\mu F_1 + {i\sigma_{\mu\nu} q^\nu \over 2
m_{\mbox{\tiny${1\over 2}^+$}}} F_2 \Bigg]u(p)~, \\
\label{ebozo08}
\la \mbox{\small ${1\over 2}^-$} (p^{\prime}) \vel j_\mu^{el} \ver
\mbox{\small ${1\over 2}^+$} (p) \ra =         
\bar{u}(p^\prime) \Bigg[\Bigg(\gamma_\mu  - {\rlap/{q} q_\mu \over
q^2}\Bigg) F_1^\ast + {i\sigma_{\mu\nu} q^\nu \over m_{\mbox{\tiny${1\over
2}^+$}}+m_{\mbox{\tiny${1\over 2}^-$}}}  F_2^\ast \Bigg] \gamma_5 u(p)~,
\eea
where $\ve \mbox{\small ${1\over 2}^+$} (p) \ra$ state corresponds
to the one nucleon state.

The matrix elements which describe the contributions of the spin-3/2 states
are defined as:
\bea
\label{ebozo09}
\la 0 \vel j_\alpha \ver
\mbox{\small ${3\over 2}^+$} (p^\prime) \ra \es \lambda_+ u_\alpha
(p^\prime)~, \nnb \\
\la 0 \vel j_\alpha \ver                                          
\mbox{\small ${3\over 2}^-$} (p^\prime) \ra \es \lambda_- \gamma_5 u_\alpha
(p^\prime)~, \\
\label{ebozo10}
\la \mbox{\small ${3\over 2}^+$} (p^\prime) \vel j_\mu^{el} \ver
\mbox{\small ${1\over 2}^+$} (p) \ra \es 
\bar{u}_\beta (p^\prime) \Big\{ G_1^+ (q_\beta \gamma_\mu - \rlap/{q}
g_{\beta\mu} ) + G_2^+ [ q_\beta {\cal P}_\mu - (q {\cal P}) g_{\beta\mu}] \nnb \\
\ar G_3^+ (q_\beta q_\mu - q^2 g_{\beta\mu} ) \Big\} \gamma_5 u(p)~,\\
\label{ebozo11}
\la \mbox{\small ${3\over 2}^-$} (p^\prime) \vel j_\mu^{el} \ver
\mbox{\small ${1\over 2}^+$} (p) \ra \es 
\bar{u}_\beta (p^\prime) \Big\{ G_1^- (q_\beta \gamma_\mu - \rlap/{q}
g_{\beta\mu} ) + G_2^- [ q_\beta {\cal P}_\mu - (q {\cal P}) g_{\beta\mu}] \nnb \\
\ar G_3^- (q_\beta q_\mu - q^2 g_{\beta\mu} ) \Big\}u(p)~,
\eea 
where ${\cal P}={1\over 2} (p+p^\prime)$; $q=p^\prime-p$; and $G_i^\pm$ are
the form form factors. Using the definitions of the residues and form
factors given in Eqs. (\ref{ebozo09})-(\ref{ebozo11}), and summing over
spins of spin-3/2 states given as:
\bea
\label{nolabel01}
\sum u_\alpha (p^\prime) \bar{u}_\beta (p^\prime) \es (\rlap/{p}^\prime + 
m_{\mbox{\tiny${3\over 2}^\pm$}}) \Bigg[ g_{\alpha\beta} - {1\over 3}
\gamma_\alpha \gamma_\beta - {2\over 3}{p_\alpha^\prime p_\beta^\prime \over
m_{\mbox{\tiny${3\over 2}^\pm$}}^2} + {1\over 3} {p_\alpha^\prime
\gamma_\beta - p_\beta^\prime \gamma_\alpha \over m_{\mbox{\tiny${3\over
2}^\pm$}}} \Bigg]~,\nnb
\eea
for the phenomenological part we obtain,
\bea
\label{ebozo12}  
\Pi_{\alpha\mu} \es - {\lambda_- \over m_{\mbox{\tiny${3\over
2}^-$}}-p^{\prime 2}} \gamma_5
(\rlap/{p}^\prime + m_{\mbox{\tiny${3\over 2}^-$}}) \Bigg[
g_{\alpha\beta} - {1\over 3}
\gamma_\alpha \gamma_\beta - {2\over 3}{p_\alpha^\prime p_\beta^\prime \over
m_{\mbox{\tiny${3\over 2}^\pm$}}^2} + {1\over 3} {p_\alpha^\prime
\gamma_\beta - p_\beta^\prime \gamma_\alpha \over m_{\mbox{\tiny${3\over
2}^\pm$}}} \Bigg] \nnb \\
\cp \Big\{ G_1^- (q_\beta \gamma_\mu - \rlap/{q}
g_{\beta\mu} ) + G_2^- [ q_\beta {\cal P}_\mu - (q {\cal P}) g_{\beta\mu}]
+ G_3^- (q_\beta q_\mu - q^2 g_{\beta\mu} ) \Big\}u(p) \nnb \\
\ek {\lambda_+ \over m_{\mbox{\tiny${3\over
2}^+$}}-p^{\prime 2}}
(\rlap/{p}^\prime + m_{\mbox{\tiny${3\over 2}^+$}}) \Bigg[
g_{\alpha\beta} - {1\over 3}
\gamma_\alpha \gamma_\beta - {2\over 3}{p_\alpha^\prime p_\beta^\prime \over
m_{\mbox{\tiny${3\over 2}^\pm$}}^2} + {1\over 3} {p_\alpha^\prime
\gamma_\beta - p_\beta^\prime \gamma_\alpha \over m_{\mbox{\tiny${3\over
2}^\pm$}}} \Bigg] \nnb \\
\cp \Big\{ G_1^+ (q_\beta \gamma_\mu - \rlap/{q}
g_{\beta\mu} ) + G_2^+ [ q_\beta {\cal P}_\mu - (q {\cal P}) g_{\beta\mu}]
+ G_3^+ (q_\beta q_\mu - q^2 g_{\beta\mu} ) \Big\} \gamma_5 u(p) \nnb \\
\ek {B_+ \over m_{\mbox{\tiny${1\over 2}^+$}}^2 -p^{\prime 2}}
\Bigg( \gamma_\alpha - {4\over m_{\mbox{\tiny${1\over 2}^+$}}}
p_\alpha^\prime \Bigg) (\rlap/{p}^\prime + m_{\mbox{\tiny${1\over 2}^+$}})
\Bigg[ F_1\gamma_\mu - {F_2 \over 4 m_{\mbox{\tiny${1\over 2}^+$}}}
(\gamma_\mu \rlap/{q} - \rlap/{q} \gamma_\mu) \Bigg] u(p) \nnb \\ 
\ek {B_- \over m_{\mbox{\tiny${1\over 2}^-$}}^2 -p^{\prime 2}} \gamma_5
\Bigg( \gamma_\alpha - {4\over m_{\mbox{\tiny${1\over 2}^-$}}}
p_\alpha^\prime \Bigg) (\rlap/{p}^\prime + m_{\mbox{\tiny${1\over 2}^-$}})
\Bigg[ F_1^\ast \Bigg(\gamma_\mu - {\rlap/{q} \gamma_\mu \over q^2} \Bigg) \nnb \\
\ek {F_2^\ast \over 2 (m_{\mbox{\tiny${1\over 2}^+$}} +
m_{\mbox{\tiny${1\over 2}^-$}})}
(\gamma_\mu \rlap/{q} - \rlap/{q} \gamma_\mu)
\Bigg] \gamma_5 u(p) 
\eea
One can easily see from Eq. (\ref{ebozo12}) that, in the first two terms
which describe the contributions of the spin-3/2 states, all structures,
except $g_{\alpha\beta}$ contain also contributions coming from spin-1/2
states which should all be removed. We see from Eq. (\ref{ebozo12}) that the
terms with $\gamma_\alpha$ at left or proportional to $p_\alpha^\prime$, as
well as $\gamma_\mu$ at right or terms proportional to $p_\mu^\prime$,
describe the contributions of the spin-1/2 states. Consider the terms
multiplied by the structure $\rlap/{p}^\prime \gamma_\alpha$ which can be
written as $\rlap/{p}^\prime \gamma_\alpha = 2 p_\alpha^\prime -
\gamma_\alpha \rlap/{p}^\prime$. Both terms on the right in this identity
contain the contributions coming from spin-1/2 states. As a result one
can conclude that, in Eq. (\ref{ebozo12}) only the terms that are proportional
to the $g_{\alpha\beta}$ structure describe the contributions of spin-3/2 states.  
So, the hadronic part of the correlation function that involves
contributions from both negative and positive parity spin-3/2 states only
can be written as:
\bea
\label{ebozo13}
\Pi_{\alpha\mu} \es {\lambda_- \over m_{\mbox{\tiny${3\over
2}^-$}}-p^{\prime 2}} (\rlap/{p}^\prime + m_{\mbox{\tiny${3\over 2}^-$}})
\Bigg[G_1^- (- q_\alpha \gamma_\mu + \rlap/{q} g_{\alpha\mu})
+ G_2^- [ q_\alpha {\cal P}_\mu - (q {\cal P}) g_{\alpha\mu}] \nnb \\
\ar G_3^- (q_\alpha q_\mu - q^2 g_{\alpha\mu} ) \Big] \gamma_5 u(p)
\nnb \\
\ek {\lambda_+ \over m_{\mbox{\tiny${3\over 2}^+$}}-p^{\prime 2}}
(\rlap/{p}^\prime + m_{\mbox{\tiny${3\over 2}^+$}})   
\Bigg[G_1^+ (q_\alpha \gamma_\mu - \rlap/{q} g_{\alpha\mu})
+ G_2^+ [ q_\alpha {\cal P}_\mu - (q {\cal P}) g_{\alpha\mu}] \nnb \\
\ar G_3^+ (q_\alpha q_\mu - q^2 g_{\alpha\mu} ) \Big] \gamma_5 u(p)~.
\eea
We see from this equation that the correlation function contains numerous
Lorentz structures all of which are not independent of each other. In order to 
allocate independent structures ordering of the Dirac matrices is
implemented, from which we choose the ordering $\gamma_\alpha \rlap/{p}
\rlap/{q} \gamma_\mu \gamma_5$. The correlation function (\ref{ebozo01})
can be decomposed in terms of the independent invariant amplitudes as
follows:
\bea
\label{ebozo14}
\Pi_{\alpha\mu} \es \Pi_1 [(p+q)^2,q^2] \rlap/{q} \gamma_\mu \gamma_5
q_\alpha + \Pi_2 [(p+q)^2,q^2] \gamma_\mu \gamma_5 q_\alpha +
\Pi_3 [(p+q)^2,q^2] \rlap/{q} \gamma_5 p_\mu q_\alpha \nnb \\
\ar \Pi_4 [(p+q)^2,q^2] \gamma_5 p_\mu q_\alpha + 
\Pi_5 [(p+q)^2,q^2] \rlap/{q} \gamma_5 q_\alpha q_\mu+
\Pi_6 [(p+q)^2,q^2] \gamma_5 q_\alpha q_\mu \nnb \\
\ar \mbox{other invariant functions.}
\eea
Comparing Eqs. (\ref{ebozo13}) and (\ref{ebozo14}), we get the following set
of equations from which we can determine the form factors $G_i^\pm~(i=1,2,3)$.
\bea
\label{ebozo15}
- {\lambda_- \over m_{\mbox{\tiny${3\over 2}^-$}}-p^{\prime 2}} G_1^- -
{\lambda_+ \over m_{\mbox{\tiny${3\over 2}^+$}}-p^{\prime 2}} G_1^+
\es \Pi_1\nnb \\
{\lambda_- m_{\mbox{\tiny${3\over 2}^-$}} \over
m_{\mbox{\tiny${3\over 2}^-$}}-p^{\prime 2}} G_1^- - 
{\lambda_+ m_{\mbox{\tiny${3\over 2}^+$}}\over
m_{\mbox{\tiny${3\over 2}^+$}}-p^{\prime 2}} G_1^+
\es \Pi_2\nnb \\
{\lambda_- \over m_{\mbox{\tiny${3\over 2}^-$}}-p^{\prime 2}} G_2^- -
{\lambda_+ \over m_{\mbox{\tiny${3\over 2}^+$}}-p^{\prime 2}} G_2^+
\es \Pi_3\nnb \\
- {\lambda_- m_{\mbox{\tiny${3\over 2}^-$}} \over
m_{\mbox{\tiny${3\over 2}^-$}}-p^{\prime 2}} G_2^- - 
{\lambda_+ m_{\mbox{\tiny${3\over 2}^+$}}\over
m_{\mbox{\tiny${3\over 2}^+$}}-p^{\prime 2}} G_2^+
\es \Pi_4\nnb \\
{\lambda_- \over m_{\mbox{\tiny${3\over 2}^-$}}-p^{\prime 2}}
\Bigg({G_2^-\over 2} - G_3^-\Bigg)  -
{\lambda_+ \over m_{\mbox{\tiny${3\over 2}^+$}}-p^{\prime 2}}
\Bigg( {G_2^- \over 2} - G_3^+\Bigg)
\es \Pi_5\nnb \\
- {\lambda_- m_{\mbox{\tiny${3\over 2}^-$}} \over
m_{\mbox{\tiny${3\over 2}^-$}}-p^{\prime 2}}
\Bigg({G_2^-\over 2} - G_3^-\Bigg)  -
{\lambda_+ m_{\mbox{\tiny${3\over 2}^+$}} \over
m_{\mbox{\tiny${3\over 2}^+$}}-p^{\prime 2}}
\Bigg( {G_2^+ \over 2} - G_3^+\Bigg)
\es \Pi_6~.
\eea
From these equations we finally obtain the following expressions for the
transition form factors $G_i^-~(i=1,2,3)$:
\bea
\label{ebozo16}
{\lambda_- (m_{\mbox{\tiny${3\over 2}^-$}}+ m_{\mbox{\tiny${3\over 2}^+$}})
\over m_{\mbox{\tiny${3\over 2}^-$}}-p^{\prime 2}} G_1^- \es
- m_{\mbox{\tiny${3\over 2}^+$}} \Pi_1 + \Pi_2 \nnb \\
{\lambda_- (m_{\mbox{\tiny${3\over 2}^-$}}+ m_{\mbox{\tiny${3\over 2}^+$}})
\over m_{\mbox{\tiny${3\over 2}^-$}}-p^{\prime 2}} G_2^- \es 
m_{\mbox{\tiny${3\over 2}^+$}} \Pi_3 - \Pi_4 \nnb \\
{\lambda_- (m_{\mbox{\tiny${3\over 2}^-$}}+ m_{\mbox{\tiny${3\over 2}^+$}})
\over m_{\mbox{\tiny${3\over 2}^-$}}-p^{\prime 2}}
\Bigg( {G_2^- \over 2} - G_3^-\Bigg)
\es m_{\mbox{\tiny${3\over 2}^+$}} \Pi_5 - \Pi_6~.
\eea

Using the quark hadron duality ansatz, 
we next calculate the invariant functions $\Pi_i$ from the QCD side.
In order to justify the operator product expansion (OPE) near the light
cone $x^2 \simeq 0$, the external momenta $(p+q)^2$ and $q^2$ are taken to
be space-like. The result of OPE is obtained in terms of the nucleon
distribution amplitudes (DAs) of growing twist. The nucleon DAs are the key
ingredient of the LCSR and they are calculated in \cite{Rbozo09} up to
twist-6. Having the explicit forms of the nucleon DAs, the the invariant
amplitudes $\Pi_i~(i=1,\cdots,6)$ can be written in the following form:
\bea
\label{ebozo17}  
\Pi_i[(p+q)^2,q^2] = \sum_{i=1}^3 \int_0^1
{\rho_{i,n} [x,q^2,(p+q)^2] \over [(q+px)^2 ]^n }~,
\eea
where $i=1,\cdots,6$; $n=1,2,3$. The expressions of $\rho_{i,n}$ for the 
form factors $G_1(Q^2)$, $G_2(Q^2)$ and ${\ds G_2(Q^2) \over \ds 2}
- G_3(Q^2)$ are presented in Appendix A.

Substituting the results for $\Pi_i$ given in Eq. (\ref{ebozo17}) into Eq.
(\ref{ebozo16}), and performing Borel transformation with respect to
$(p+q)^2$ in order to suppress contributions of the higher states and
continuum, we obtain the following sum rules for the form factors $G_1^-(Q^2)$,
$G_2^-(Q^2)$ and  ${\ds G_2^-(Q^2) \over \ds 2}-G_3^-(Q^2)$:
\bea
\label{ebozo18}  
\lambda_- (m_{\mbox{\tiny${3\over 2}^-$}} + m_{\mbox{\tiny${3\over 2}^+$}})
G_1^-(Q^2) e^{-m_{\mbox{\tiny${3\over 2}^-$}}/M^2} \es
- m_{\mbox{\tiny${3\over 2}^+$}} I_1(Q^2,M^2,s_0) + I_2(Q^2,M^2,s_0)~,
\nnb \\
\lambda_- (m_{\mbox{\tiny${3\over 2}^-$}} + m_{\mbox{\tiny${3\over 2}^+$}})
G_2^-(Q^2) e^{-m_{\mbox{\tiny${3\over 2}^-$}}/M^2} \es
m_{\mbox{\tiny${3\over 2}^+$}} I_3(Q^2,M^2,s_0) - I_4(Q^2,M^2,s_0)~,
\nnb \\
\lambda_- (m_{\mbox{\tiny${3\over 2}^-$}} + m_{\mbox{\tiny${3\over 2}^+$}})
\Bigg[{G_2^-(Q^2)\over 2} - G_3^-(Q^2) \Bigg]
e^{-m_{\mbox{\tiny${3\over 2}^-$}}/M^2} \es
m_{\mbox{\tiny${3\over 2}^+$}} I_5(Q^2,M^2,s_0) \nnb \\
\ek I_6(Q^2,M^2,s_0)~,
\eea
where
\bea
\label{ebozo19}
I_i(Q^2,M^2,s_0) \es
\int_{x_0}^1 dx \Bigg[
- {\rho_{i,1}(x)\over x} + {\rho_{i,2}(x) \over x^2 M^2} -
- {\rho_{i,3}(x) \over 2 x^3 M^4} \Bigg] e^{-s(x)/M^2}\nnb \\
\ar \Bigg[ {\rho_{i,2}(x_0) \over Q^2 + x_0^2 m_B^2} -
{1\over 2 x_0}
{\rho_{i,3}(x_0) \over (Q^2 + x_0^2 m_B^2) M^2} \nnb \\
\ar {1\over 2} {x_0^2 \over (Q^2 + x_0^2 m_B^2)} \Bigg(
{d\over dx_0} {\rho_{i,3}(x_0) \over x_0 (Q^2 + x_0^2 m_B^2)
M^2} \Bigg) \Bigg]e^{-s_0/ M^2}
\Bigg\}~,
\eea
where $M^2$ is the Borel mass square, $s_0$ is the continuum threshold and
\bea
\label{nolabel02}
s(x) = {\bar{x} Q^2 + x \bar{x} m_B^2 \over x}~, \nnb
\eea
$\bar{x} = 1-x$, and $x_0$ is the solution of the equation $s(x) = s_0$.

It follows from Eq. (\ref{ebozo19}) that the residue of the negative parity
spin-3/2 $N^\ast(1520)$ baryon is needed. The decay constant $\lambda_-$
of the negative parity $N^\ast(1520)$ baryon can be obtained from the
following two-point correlation function
\bea
\label{ebozo20}
\Pi_{\alpha\beta} (p^2) \es i \int d^4x e^{ipx} \lla 0 \vel \eta_\alpha(x)
\bar{\eta}_\beta (0) \ver 0 \rra \nnb \\
\es \Pi_1(p^2) g_{\alpha\beta} + \Pi_2(p^2) p_\alpha p_\beta + \Pi_3(p^2)
\gamma_\alpha \gamma_\beta + \Pi_4(p^2) \gamma_\alpha p_\beta + \Pi_5(p^2)
\gamma_\beta p_\alpha~.
\eea

As we have already noted, the structures $p_\alpha p_\beta$, $\gamma_\alpha
\gamma_\beta$, $\gamma_\alpha p_\beta$ and $\gamma_\beta p_\alpha$ contain
contributions from spin-1/2 states. Therefore, in order to determine the
decay constant of the negative parity spin-3/2 baryon, the above-mentioned
structures should be omitted. The only structure which contains
contributions from spin-3/2 states with both parity is $g_{\mu\nu}$.
Retaining only this structure, and saturating (\ref{ebozo20}) with spin-3/2
states, from the physical part of the correlation function we get,
\bea
\label{ebozo21}
{\vel \lambda_+\ver^2 (\not\!p + m_{\mbox{\tiny${3\over 2}^+$}}) \over 
m_{\mbox{\tiny${3\over 2}^+$}}-p^2} +
{\vel \lambda_-\ver^2 (\not\!p - m_{\mbox{\tiny${3\over 2}^-$}}) \over 
m_{\mbox{\tiny${3\over 2}^-$}}-p^2} + \cdots
\eea
Denoting the spectral densities $\rho_1(s)$ and
$\rho_2(s)$ corresponding to the structures $\not\!p g_{\alpha\beta}$ and
$g_{\alpha\beta}$, respectively, and performing Borel transformation over
$-p^2$ we get,
\bea
\label{ebozo22}
\vel \lambda_+\ver^2 e^{-m_{\mbox{\tiny${3\over 2}^+$}}/M^2} +
\vel \lambda_-\ver^2 e^{-m_{\mbox{\tiny${3\over 2}^-$}}/M^2} \es \int_0^{s_0}
ds e^{-s/M^2} \rho_1(s)~, \nnb \\
m_{\mbox{\tiny${3\over 2}^+$}} \vel \lambda_+\ver^2 e^{-m_{\mbox{\tiny${3\over
2}^+$}}/M^2} - \vel \lambda_-\ver^2 m_{\mbox{\tiny${3\over 2}^-$}}
e^{-m_{\mbox{\tiny${3\over 2}^-$}}/M^2} \es \int_0^{s_0} 
ds e^{-s/M^2} \rho_2(s)~.
\eea
Solving these equations for the mass and decay constant of the negative
parity spin-3/2 $N^\ast (1520)$ baryon, we get
\bea
\label{ebozo23}
m_{\mbox{\tiny${3\over 2}^-$}} \es {\ds \int_0^{s_0} ds e^{-s/M^2} s
\Big[m_{\mbox{\tiny${3\over 2}^+$}} \rho_1(s) - \rho_2(s) \Big] \over
\ds \int_0^{s_0} ds e^{-s/M^2}
\Big[m_{\mbox{\tiny${3\over 2}^+$}} \rho_1(s) - \rho_2(s) \Big]}~ \\
\label{ebozo24}
\vel \lambda_-\ver^2 \es {e^{m_{\mbox{\tiny${3\over 2}^-$}}/M^2} \over
(m_{\mbox{\tiny${3\over 2}^-$}} + m_{\mbox{\tiny${3\over 2}^+$}})}
\int_0^{s_0} ds e^{-s/M^2} \Big[m_{\mbox{\tiny${3\over 2}^+$}} \rho_1(s) -
\rho_2(s) \Big]~.
\eea
In order to calculate the the spectral densities $\rho_1$ and
$\rho_2$, we need to obtain the expressions $\Pi_1$ and $\Pi_2$
corresponding to the structures $\not\!p g_{\alpha\beta}$ and
$g_{\alpha\beta}$, respectively. The explicit forms of $\Pi_1$ and $\Pi_2$
after the Borel transformation is performed, are given in Appendix B (see
also \cite{Rbozo10}). The
spectral densities $\rho_1(s)$ and $\rho_2(s)$ can be obtained after
applying the Fourier transformation and performing continuum subtraction to
these expressions.    
 
In general, the electromagnetic properties of hadrons can also be described
in terms of the helicity amplitudes or multipole electromagnetic form
factors instead of the form factors $G_1$, $G_2$ and $G_3$. The multipole
electromagnetic form factors in our case are the magnetic dipole $G_M$,
electric quadrupole $G_E$ and Coulomb quadrupole $G_C$. These form factors
are defined as (see for example \cite{Rbozo06}):
\bea
\label{ebozo25}
G_M \es -{1\over \sqrt{6}} {m_N\over m_{\mbox{\tiny${3\over 2}^-$}} - m_N}
\Bigg[ (m_{\mbox{\tiny${3\over 2}^-$}} - m_N)^2 + Q^2 \Bigg] {G_1\over
m_{\mbox{\tiny${3\over 2}^-$}} }~, \\
\label{ebozo26}
G_E \es -{1\over \sqrt{6}} {m_N\over m_{\mbox{\tiny${3\over 2}^-$}} - m_N}
\Bigg[ (m_{\mbox{\tiny${3\over
2}^-$}}^2  - m_N^2 - Q^2 ) {G_1 \over m_{\mbox{\tiny${3\over 2}^-$}} }  + (m_{\mbox{\tiny${3\over
2}^-$}}^2  - m_N^2) G_2 - 2 Q^2 G_3 \Bigg] \\
\label{ebozo27}
G_C \es -{1\over \sqrt{6}} {m_N\over m_{\mbox{\tiny${3\over 2}^-$}} - m_N}
\Bigg[ 4 m_{\mbox{\tiny${3\over 2}^-$}} G_1 + (3 m_{\mbox{\tiny${3\over
2}^-$}}^2  + m_N^2 + Q^2 ) G_2 \nnb \\
\ar 2 ( m_{\mbox{\tiny${3\over 2}^-$}}^2  - m_N^2 - Q^2 ) G_3 \Bigg]
\eea
   
The relations among helicity amplitudes and multipole form factors are given
as follows:

\bea
\label{ebozo28}
A_{1/2} \es {1\over 4 C} (3 G_M - G_E) \\
\label{ebozo29}
A_{3/2} \es {\sqrt{3} \over 4 C} (G_M + G_E) \\
\label{ebozo30}
S_{1/2} \es - {\ve \vec{q} \ve e \over
4 \sqrt{2} m_{\mbox{\tiny${3\over 2}^-$}} } \sqrt{
\left[ {(m_{\mbox{\tiny${3\over 2}^-$}} + m_N)^2 + Q^2 \over
6 m_{\mbox{\tiny${3\over 2}^-$}} m_N } \right] \left[ {2 m_{\mbox{\tiny${3\over 2}^-$}}
\over  m_{\mbox{\tiny${3\over 2}^-$}}^2 - m_N^2 } \right] } g_C~,
\eea 

where
\bea
\label{nolabel03}
C \es {m_N \over e \ve \vec{q} \ve} \sqrt{
\left[ { m_N (m_{\mbox{\tiny${3\over 2}^-$}}^2 - m_N^2) \over
2 m_{\mbox{\tiny${3\over 2}^-$}}^2 } \right] \left[
{(m_{\mbox{\tiny${3\over 2}^-$}} - m_N)^2 + Q^2 \over
(m_{\mbox{\tiny${3\over 2}^-$}} - m_N)^2 } \right] }~, \nnb \\ \nnb \\
g_C \es\ 4  m_{\mbox{\tiny${3\over 2}^-$}} G_1 + (3 m_{\mbox{\tiny${3\over
2}^-$}}^2  + m_N^2 + Q^2 ) G_2
+ 2 ( m_{\mbox{\tiny${3\over 2}^-$}}^2  -
m_N^2 - Q^2 ) G_3~,~\mbox{\rm and,} \nnb \\ \nnb \\
\ve \vec{q} \ve  \es {1 \over 2 m_{\mbox{\tiny${3\over 2}^-$}} } \sqrt{ \Big[
(m_{\mbox{\tiny${3\over 2}^-$}} + m_N )^2 + Q^2 \Big]
\Big[ (m_{\mbox{\tiny${3\over 2}^-$}} -  m_N )^2 + Q^2 \Big]
}
\eea
is the magnitude of the nucleon three-momentum in the rest frame of the
$N^\ast (1520)$ baryon.

\section{Numerical analysis}

Here in this section we present the numerical results of the LCSR
prediction on the multipole form factors, as well as helicity
amplitudes for the $\gamma^\ast N \to N^\ast (1520)$ transition.
The main input parameters of this method are the distribution
amplitudes (DAs), and we use the results of \cite{Rbozo09} in the following
numerical calculations. The values of the factors appearing in the nucleon
DAs are calculated within the LCSR, and they are given as:
\bea
\label{nolabel05}
f_N \es (5.0 \pm 0.5) \times 10^{-3}~GeV^2~,\mbox{\cite{Rbozo09}}\nnb \\
\lambda_1 \es - (2.7 \pm 0.9) \times 10^{-2}~GeV^2~,\mbox{\cite{Rbozo11}}\nnb \\
\lambda_2 \es - (5.4 \pm 1.9) \times 10^{-2}~GeV^2~,\mbox{\cite{Rbozo11}}\nnb
\eea
and $A_1^u = 0.13$, $V_1^d = 0.3$, $f_1^d = 0.33$, and $f_2^d = 0.25$
\cite{Rbozo11}. In calculation of the transition form factors responsible
for the $\gamma^\ast N \to N^\ast (1520)$ transition masses and decay
constants of the negative parity baryons, which are determined from the
analysis of the two-point sum rules (see Eq. (\ref{ebozo22})). These sum
rules have two auxiliary parameters, namely, the Borel mass parameter $M^2$ and
the continuum threshold $s_0$. The working window for the Borel mass
parameter is determined from the condition that the nonperturbative and
continuum contribution should be less than, say, half of the perturbative
contributions. The value of the continuum threshold is determined from the
condition that, the mass sum rules reproduces the experimental value of the
$N^\ast(1520)$ mass with $10\%$ accuracy. 

The analysis of the mass sum rules shows that the criteria for the Borel
mass parameter which has been noted above, is fulfilled in the region $1.5
\le M^2 \le 3.0~GeV^2$. We also observe that if the continuum threshold is
chosen in the region $s_0=(4.5 \pm 0.5)~GeV^2$, sum rules reproduce the mass
of the $N^\ast(1520)$ baryon with $10\%$ accuracy, as the result of which we
get $m_-=(5.4 \pm0.1)~GeV$.
In further analysis we shall use this value of $m_-$,
and the above-mentioned regions of $M^2$ and $s_0$.

In Figs. 1, 2 and 3 we present the dependencies of the helicity amplitudes
$A_{1/2}$, $A_{3/2}$ and  $S_{1/2}$ on $Q^2$, respectively,
at several fixed values of the
Borel parameter $M^2$ and at $s_0=4.0~GeV^2$. In the numerical analysis we
take into account the errors in the input parameters and the graphs are
presented at the minimum values of the input parameters. In order to keep the higher twist
and higher states contributions under control, $Q^2$ is running in the
region $1.0 \le Q^2 \le 5.0~GeV^2$. From all these we figures we see that,
the trend of behavior for all three figures are the same, i.e., all three
amplitudes tend to zero with increasing $Q^2$. Moreover, $A_{3/2}$ is
always positive, while $A_{1/2}$ and $S_{1/2}$ are alway negative for all
values of $Q^2$. If the maximum values of the input parameters are taken
into account, the
values of the helicity amplitudes increase slightly in modulo. 
Among all these three amplitudes, $A_{1/2}$ is less
sensitive to the errors in the  input parameters.

When we compare our results with the predictions of the spectator quark
model \cite{Rbozo06}, we observe that our results on the value of $A_{1/2}$
are approximately the same as with the prediction of \cite{Rbozo06}, and very
close to the experimental result \cite{Rbozo07}.
In the case of $S_{1/2}$ amplitude, our results are slightly smaller 
compared those
given in \cite{Rbozo06}. This result demonstrates the necessity for more
precise determination of the input parameters. In the case for the $A_{3/2}$
amplitude, our results are very close to the ones given in \cite{Rbozo06}
within the limits of errors, as well as to the experimental data given in
\cite{Rbozo07}.

Our prediction on the form factors and helicity amplitudes can be refined
further by taking the radiative corrections into account for the
distribution amplitudes. The first calculation along this line has been
carried out in \cite{Rbozo12}.

In summary, we study the $\gamma^\ast N \to N^\ast (1520)$ transition within
the LCSR method. We derive the sum rules for the magnetic
dipole $G_M$, electric quadrupole $G_E$, and Coulomb quadrupole $G_C$ form
factors, and consequently for the helicity amplitudes $A_{1/2}$,
$A_{3/2}$ and $S_{1/2}$. The $Q^2$ dependence of these helicity amplitudes
are investigated. Finally, we compare our results with the predictions of
the covariant spectator quark model, and with the existing experimental
data.     

\newpage

\section*{Appendix A}  
\setcounter{equation}{0}

In this appendix we present the expressions for the functions
$\rho_2$, $\rho_4$ and $\rho_6$
which appear in the sum rules for $G_1(Q^2)$,
$G_2(Q^2)$, and $\ds{G_2(Q^2)\over 2} - G_3(Q^2)$.

\section*{For the form factor $G_1$}

\bea
%
\rho_6 (x)\es
-32 (1-x) m_N^2 m_{\mbox{\tiny${3\over 2}^+$}} (x^2 m_N^2 +Q^2) \,
(e_{q_1} \, \check{\!\check{B}}_6 + e_{q_2} 
\; \widetilde{\!\widetilde{B}}_6)~, \nnb \\ \nnb \\
\rho_4 (x) \es
8 e_{q_1} m_N^2 \Big[ 2 (1-x) m_N (\, \check{\!\check{B}}_8                    
+ \, \check{\!\check{C}}_6 + \, \check{\!\check{D}}_6)
- (1-2 x) m_{\mbox{\tiny${3\over 2}^+$}} \, \check{\!\check{B}}_6 \Big] \nnb \\
\ar 8 e_{q_2} m_N^2 \Big[ 2 (1-x) m_N (\; \widetilde{\!\widetilde{B}}_8
+ \; \widetilde{\!\widetilde{C}}_6 + \; \widetilde{\!\widetilde{D}}_6)
- (1-2 x) m_{\mbox{\tiny${3\over 2}^+$}} \; \widetilde{\!\widetilde{B}}_6 \Big] \nnb \\
\ar 16 e_{q_3} m_N^2 \Big[ 2 (1-x) m_N  \, \widehat{\!\widehat{B}}_8    
- x m_{\mbox{\tiny${3\over 2}^+$}} \, \widehat{\!\widehat{B}}_6 \Big] \nnb \\
\ek 8 e_{q_1} {1-x\over x} m_N \Big\{
(x^2 m_N^2 +Q^2) \check{B}_2 + \Big[x^2 m_N^2 - (1-2 x) Q^2 \Big] \check{B}_4 \nnb \\
\ar 2 x \Big[ (x m_N^2 + Q^2) (\check{C}_2 + \check{D}_2)
- x m_N m_{\mbox{\tiny${3\over 2}^+$}} (\check{B}_5 + \check{C}_5 -
\check{D}_5 - \check{P}_2 - \check{S}_2) \Big] \Big\} \nnb \\
%
%
\ek 8 e_{q_2} {1-x\over x} m_N \Big\{
(x^2 m_N^2 +Q^2) \widetilde{B}_2 + \Big[x^2 m_N^2 - (1-2 x) Q^2  \Big] \widetilde{B}_4 \nnb \\
\ek 2 x \Big[ (x m_N^2 + Q^2) (\widetilde{C}_2 + \widetilde{D}_2)
+ x m_N m_{\mbox{\tiny${3\over 2}^+$}} (\widetilde{B}_5 + 
\widetilde{P}_2 + \widetilde{S}_2) \Big] \Big\} \nnb \\
%
%
\ar 8 e_{q_3} (1-x) m_N \Big[ 2 (x m_N^2 + Q^2) (\widehat{B}_2 +                   
\widehat{B}_4) + x m_N m_{\mbox{\tiny${3\over 2}^+$}} ( \widehat{C}_4 - \widehat{C}_5 -
\widehat{D}_4 + \widehat{D}_5) \Big] \nnb \\
%
%
\ar 16 e_{q_1} m_N^2 m_{\mbox{\tiny${3\over 2}^+$}} \int_0^{\bar{x}} \,dx_3 \Big[A_1^M -
V_1^M - (1+x) T_1^M \Big](x,1-x-x_3,x_3) \nnb \\
\ek 16 e_{q_2} m_N^2 m_{\mbox{\tiny${3\over 2}^+$}} \int_0^{\bar{x}} \,dx_1 \Big[A_1^M -
V_1^M + (1+x) T_1^M\Big](x_1,x,1-x_1-x) \nnb \\
\ar 16 e_{q_3} m_N^2 m_{\mbox{\tiny${3\over 2}^+$}} \int_0^{\bar{x}}dx_1\, \Big[
x (A_1^M - V_1^M) + 2 T_1^M\Big]  (x_1,1-x_1-x,x) \nnb \\ \nnb \\
\rho_2 (x) \es
- 8 e_{q_1} {m_N\over x} \Big[\check{B}_2 - (1- 2 x) \check{B}_4 +  
2 x (\check{C}_2 + \check{D}_2) \Big] \nnb \\
\ek 8 e_{q_2} {m_N\over x} \Big[\widetilde{B}_2 - (1- 2 x) 
\widetilde{B}_4 \Big] \nnb \\
\ek 16 e_{q_1} \int_0^{\bar{x}} \,dx_3 \Big\{
(1-x) m_N (A_3-P_1+S_1+V_3) \nnb \\
\ek m_{\mbox{\tiny${3\over 2}^+$}} \Big[ A_1- (1+x) T_1 -V_1 \Big]
\Big\} (x,1-x-x_3,x_3) \nnb \\\nnb \\
%
%
\ek 16 e_{q_2} \int_0^{\bar{x}} \,dx_1 \Big\{
(1-x) m_N (P_1-S_1) \nnb \\
\ar m_{\mbox{\tiny${3\over 2}^+$}} \Big[A_1+ (1+x) T_1 - V_1 \Big] 
\Big\} (x_1,x,1-x_1-x) \nnb \\\nnb \\
%
%
\ar 16 e_{q_3} \int_0^{\bar{x}}dx_1\, \Big\{
(1-x) m_N (A_3+V_3) \nnb \\
\ar m_{\mbox{\tiny${3\over 2}^+$}} 
\Big[ x (A_1-V_1) + 2 T_1 \Big] \Big\} (x_1,1-x_1-x,x)~.
%
%
\eea

\section*{For the form factor $G_2$}
\bea
\rho_6(x) \es
-64 e_{q_1} (1-x) m_N^2 \Big\{                                      
\Big[x^2 m_N^2 - (1-2 x) Q^2 \Big] \, \check{\!\check{B}}_6
+ 2 x^2 m_N m_{\mbox{\tiny${3\over 2}^+$}} 
(2  \, \check{\!\check{B}}_8 +  \,             
\check{\!\check{C}}_6 +  \, \check{\!\check{D}}_6 ) \Big\} \nnb \\
\ek 64 e_{q_2} (1-x) m_N^2 \Big\{
%
%
\Big[x^2 m_N^2 - (1- 2 x) Q^2 \Big]  \; \widetilde{\!\widetilde{B}}_6
%
+ 2 x^2 m_N m_{\mbox{\tiny${3\over 2}^+$}} ( 2  \;               
\widetilde{\!\widetilde{B}}_8 +  \; \widetilde{\!\widetilde{C}}_6
+  \; \widetilde{\!\widetilde{D}}_6) \Big\} \nnb \\
\ek 128 e_{q_3} x (1-x) m_N^2 
\Big\{ 
(x m_N^2 + Q^2) \, \widehat{\!\widehat{B}}_6
+ x m_N m_{\mbox{\tiny${3\over 2}^+$}} ( 2  \,
\widehat{\!\widehat{B}}_8 +  \, \widehat{\!\widehat{C}}_6
+ \, \widehat{\!\widehat{D}}_6) \Big\} \nnb \\ \nnb \\
\rho_4 (x) \es
16 e_{q_1} (3-4 x) m_N^2 \, \check{\!\check{B}}_6 \nnb \\
\ar 16 e_{q_2} (3-4 x) m_N^2 \; \widetilde{\!\widetilde{B}}_6  \nnb \\
\ar 16 e_{q_1} x m_N  \Big\{ (1-x) m_N \Big[2 \check{B}_5
+ \check{C}_4 - \check{C}_5 - \check{D}_4 + \check{D}_5 
+ 2 (\check{E}_1 + \check{H}_1) \Big] \nnb \\
\ar  2 m_{\mbox{\tiny${3\over 2}^+$}} \Big[\check{B}_2
- (1-2 x) \check{B}_4 + x ( \check{C}_2 + \check{D}_2) \Big] \Big\} \nnb \\
%
%
\ar 16 e_{q_2} x m_N  \Big\{ (1-x) m_N \Big[2 \widetilde{B}_5
+ \widetilde{C}_4 + \widetilde{C}_5 - \widetilde{D}_4 - \widetilde{D}_5
- 2 (\widetilde{E}_1 + \widetilde{H}_1) \Big] \nnb \\
\ar  2 m_{\mbox{\tiny${3\over 2}^+$}} \Big[\widetilde{B}_2
- (1-2 x) \widetilde{B}_4 - (1-x) ( \widetilde{C}_2 + \widetilde{D}_2) \Big]
\Big\} \nnb \\
%
%
\ar 32 e_{q_3} x m_N \Big\{ (1-x) m_N (2 \widehat{B}_5 + \widehat{C}_5
- \widehat{D}_5 ) - m_{\mbox{\tiny${3\over 2}^+$}} \Big[ 2 (1-x) \widehat{B}_4 - x ( \widehat{C}_2
+ \widehat{D}_2 )\Big] \Big\} \nnb \\
%
%
\ek 32  e_{q_1} m_N^2 \int_0^{\bar{x}} \,dx_3 (1-x) \Big[A_1^M -
T_1^M - V_1^M \Big](x,1-x-x_3,x_3) \nnb \\
\ar 32 e_{q_2} m_N^2 \int_0^{\bar{x}} \,dx_1 \Big[T_1^M    
+ x (A_1^M - T_1^M - V_1^M ) \Big](x_1,x,1-x_1-x) \nnb \\
\ek 64 e_{q_3} m_N^2 \int_0^{\bar{x}}dx_1\, x \, T_1^M (x_1,1-x_1-x,x) \nnb \\ \nnb \\
\rho_2 (x) \es
- 32 e_{q_1} \int_0^{\bar{x}} \,dx_3 (1-x)       
\Big[ A_1 - T_1 - V_1 \Big] (x,1-x-x_3,x_3) \nnb \\
\ar 32 e_{q_2} \int_0^{\bar{x}} \,dx_1 \Big[ 
T_1 + x \, ( A_1 - T_1 - V_1) \Big] (x_1,x,1-x_1-x) \nnb \\
\ek 64 e_{q_3} \int_0^{\bar{x}}dx_1\, x \, T_1 (x_1,1-x_1-x,x)~.
\eea

\section*{For the form factor ${\ds {G_2\over  2}} - G_3$}

\bea
\rho_6 (x) \es
- 128 e_{q_1} (1-x)^2 m_N^2  \Big[  
( x m_N^2 + Q^2) \, \check{\!\check{B}}_6 +          
x m_N m_{\mbox{\tiny${3\over 2}^+$}} (2 \, \check{\!\check{B}}_8 + \, \check{\!\check{C}}_6
+ \, \check{\!\check{D}}_6) \Big] \nnb \\
\ek 128 e_{q_2} (1-x)^2 m_N^2  \Big[
( x m_N^2 + Q^2) \; \widetilde{\!\widetilde{B}}_6 +
x m_N m_{\mbox{\tiny${3\over 2}^+$}} (2 \; \widetilde{\!\widetilde{B}}_8 + \;
\widetilde{\!\widetilde{C}}_6
+ \; \widetilde{\!\widetilde{D}}_6) \Big] \nnb \\
\ek 128 e_{q_3} (1-x)^2 m_N^2  \Big[ 
( x m_N^2 + Q^2) \; \widehat{\!\widehat{B}}_6 +
x m_N m_{\mbox{\tiny${3\over 2}^+$}} (2 \; \widehat{\!\widehat{B}}_8 + \; \widehat{\!\widehat{C}}_6
+ \; \widehat{\!\widehat{D}}_6) \Big] \nnb \\ \nnb \\
\rho_4 (x) \es
- 32 e_{q_1} (1-x) m_N^2 \, \check{\!\check{B}}_6
- 32 e_{q_2} (1-x) m_N^2 \; \widetilde{\!\widetilde{B}}_6
- 32 e_{q_3} (1-x) m_N^2 \, \widehat{\!\widehat{B}}_6 \nnb \\
\ar 16 e_{q_1} (1-x) m_N \Big\{             
(1-x) m_N (4 \check{B}_5 + \check{C}_4 + \check{C}_5        
- \check{D}_4 - \check{D}_5) \nnb \\
\ar 2 m_{\mbox{\tiny${3\over 2}^+$}} \Big[ \check{B}_2 - (1-2 x) \check{B}_4 + x (\check{C}_2 +
\check{D}_2) \Big] \Big\} \nnb \\
\ar 16 e_{q_2} (1-x) m_N \Big\{
(1-x) m_N (4 \widetilde{B}_5 + \widetilde{C}_4 + \widetilde{C}_5
- \widetilde{D}_4 - \widetilde{D}_5) \nnb \\
\ar 2 m_{\mbox{\tiny${3\over 2}^+$}} \Big[ \widetilde{B}_2 - (1-2 x) \widetilde{B}_4 + x (\widetilde{C}_2 +
\widetilde{D}_2) \Big] \Big\} \nnb \\
\ar 16 e_{q_3} (1-x) m_N \Big\{
(1-x) m_N (4 \widehat{B}_5 + \widehat{C}_4 + \widehat{C}_5
- \widehat{D}_4 - \widehat{D}_5) \nnb \\
\ar 2 m_{\mbox{\tiny${3\over 2}^+$}} \Big[ \widehat{B}_2 - (1-2 x) \widehat{B}_4 + x (\widehat{C}_2 +
\widehat{D}_2) \Big] \Big\} \nnb \\
\ar 32 e_{q_1} m_N^2 \int_0^{\bar{x}} \,dx_3 
(1-x) \Big[A_1^M - 2  T_1^M - V_1^M \Big](x,1-x-x_3,x_3) \nnb \\
\ar 32 e_{q_2} m_N^2 \int_0^{\bar{x}} \,dx_1 
(1-x) \Big[A_1^M - 2 T_1^M - V_1^M \Big](x_1,x,1-x_1-x) \nnb \\  
\ar 32 e_{q_3} m_N^2 \int_0^{\bar{x}}dx_1\, (1-x)
\Big[A_1^M - 2 T_1^M - V_1^M \Big]  (x_1,1-x_1-x,x) \nnb \\ \nnb \\
\rho_2 (x) \es
 32 e_{q_1} \int_0^{\bar{x}} \,dx_3 (1-x)       
\Big[ A_1 - 2 T_1 -V_1 \Big] (x,1-x-x_3,x_3) \nnb \\
\ar 32 e_{q_2} \int_0^{\bar{x}} \,dx_1       
(1-x) \Big[A_1 -2 T_1 - V_1 \Big] (x_1,x,1-x_1-x) \nnb \\
\ar 32 e_{q_3} \int_0^{\bar{x}}dx_1\,
(1-x) \Big[ A_1 - 2 T_1 - V_1 \Big] (x_1,1-x_1-x,x)~.
\eea

In the above expressions for $\rho_i$ 
the functions ${\cal F}(x_i)$ are defined in the following way:

\bea
\label{nolabel06}
\check{\cal F}(x_1) \es \int_1^{x_1}\!\!dx_1^{'}\int_0^{1- x^{'}_{1}}\!\!dx_3\,
{\cal F}(x_1^{'},1-x_1^{'}-x_3,x_3)~, \nnb \\
\check{\!\!\!\;\check{\cal F}}(x_1) \es 
\int_1^{x_1}\!\!dx_1^{'}\int_1^{x^{'}_{1}}\!\!dx_1^{''}
\int_0^{1- x^{''}_{1}}\!\!dx_3\,
{\cal F}(x_1^{''},1-x_1^{''}-x_3,x_3)~, \nnb \\
\widetilde{\cal F}(x_2) \es \int_1^{x_2}\!\!dx_2^{'}\int_0^{1- x^{'}_{2}}\!\!dx_1\,
{\cal F}(x_1,x_2^{'},1-x_1-x_2^{'})~, \nnb \\
\widetilde{\!\widetilde{\cal F}}(x_2) \es 
\int_1^{x_2}\!\!dx_2^{'}\int_1^{x^{'}_{2}}\!\!dx_2^{''}
\int_0^{1- x^{''}_{2}}\!\!dx_1\,
{\cal F}(x_1,x_2^{''},1-x_1-x_2^{''})~, \nnb \\
\widehat{\cal F}(x_3) \es \int_1^{x_3}\!\!dx_3^{'}\int_0^{1- x^{'}_{3}}\!\!dx_1\,
{\cal F}(x_1,1-x_1-x_3^{'},x_3^{'})~, \nnb \\
\widehat{\!\widehat{\cal F}}(x_3) \es 
\int_1^{x_3}\!\!dx_3^{'}\int_1^{x^{'}_{3}}\!\!dx_3^{''}
\int_0^{1- x^{''}_{3}}\!\!dx_1\,
{\cal F}(x_1,1-x_1-x_3^{''},x_3^{''})~.\nnb
\eea

Definitions of the functions $B_i$, $C_i$, $D_i$, $E_1$ and $H_1$
that appear in the expressions for $\rho_i(x)$ are given as follows:

\bea
\label{nolabel07}
B_2 \es T_1+T_2-2 T_3~, \nnb \\
B_4 \es T_1-T_2-2 T_7~, \nnb \\
B_5 \es - T_1+T_5+2 T_8~, \nnb \\
B_6 \es 2 T_1-2 T_3-2 T_4+2 T_5+2 T_7+2 T_8~, \nnb \\
B_7 \es T_7-T_8~, \nnb \\
B_8 \es  -T_1+T_2+T_5-T_6+2 T_7+2T_8~, \nnb \\
C_2 \es V_1-V_2-V_3~, \nnb \\
C_4 \es -2V_1+V_3+V_4+2V_5~, \nnb \\
C_5 \es V_4-V_3~, \nnb \\
C_6 \es -V_1+V_2+V_3+V_4+V_5-V_6~, \nnb \\
D_2 \es -A_1+A_2-A_3~, \nnb \\
D_4 \es -2A_1-A_3-A_4+2A_5~, \nnb \\
D_5 \es A_3-A_4~, \nnb \\
D_6 \es A_1-A_2+A_3+A_4-A_5+A_6~, \nnb \\
E_1 \es S_1-S_2~, \nnb \\
H_1 \es P_2-P_1~. \nnb
\eea

Note that in the numerical analysis for the form factors
the masses of the light quarks are all set to zero.
\newpage

\section*{Appendix B}  
\setcounter{equation}{0}

In this appendix we present the expressions for the functions
$\Pi_2$ and $\Pi_2$ corresponding to the structures
$\not\!p g_{\alpha\beta}$ and $g_{\alpha\beta}$, respectively,
after the Borel transformation and continuum subtraction are performed.

\bea
\Pi_1\es
{3 \over \pi^4} M^6 E_2(x)  
%
- {5 \over 24 \pi^4} M^2  \GG E_0(x)
%
%
+ {28 \over 9 M^2} m_0^2 \qq^2 
%
%
%
- 16 \qq^2 \nnb \\ \nnb \\ 
%
%
\Pi_2 \es
%
+ {4 \over \pi^2} M^4 \qq E_1(x) 
%
%
- {2 \over \pi^2} m_0^2 M^2 \qq E_0(x)
%
%
%
+ {1 \over 9 \pi^2 M^2} m_0^2 \GG \qq \nnb \\
%
%
\ek {1\over 3 \pi^2} \GG \qq~,
\eea
where
\bea
\label{nolabel08}
E_n(x)=1-e^{-x}\sum_{i=0}^{n}\frac{x^i}{i!}~, \nnb
\eea
with $x = s_0/M^2$, and $s_0$ is being the continuum threshold.

\newpage

\section*{Figure captions}
{\bf Fig. (1)} The dependence of the helicity amplitude $A_{1/2}$ on $Q^2$
at $s_0=4.0~GeV^2$, and at several fixed values of the Borel mass parameter
$M^2$. \\ \\
{\bf Fig. (2)} The same as in Fig. (1), but for the helicity amplitude
$A_{3/2}$. \\ \\
{\bf Fig. (3)} The same as in Fig. (1), but for the helicity amplitude      
$S_{1/2}$.

\newpage

\begin{figure}[t]
\vskip -1.5cm
\begin{center}
\scalebox{0.785}{\includegraphics{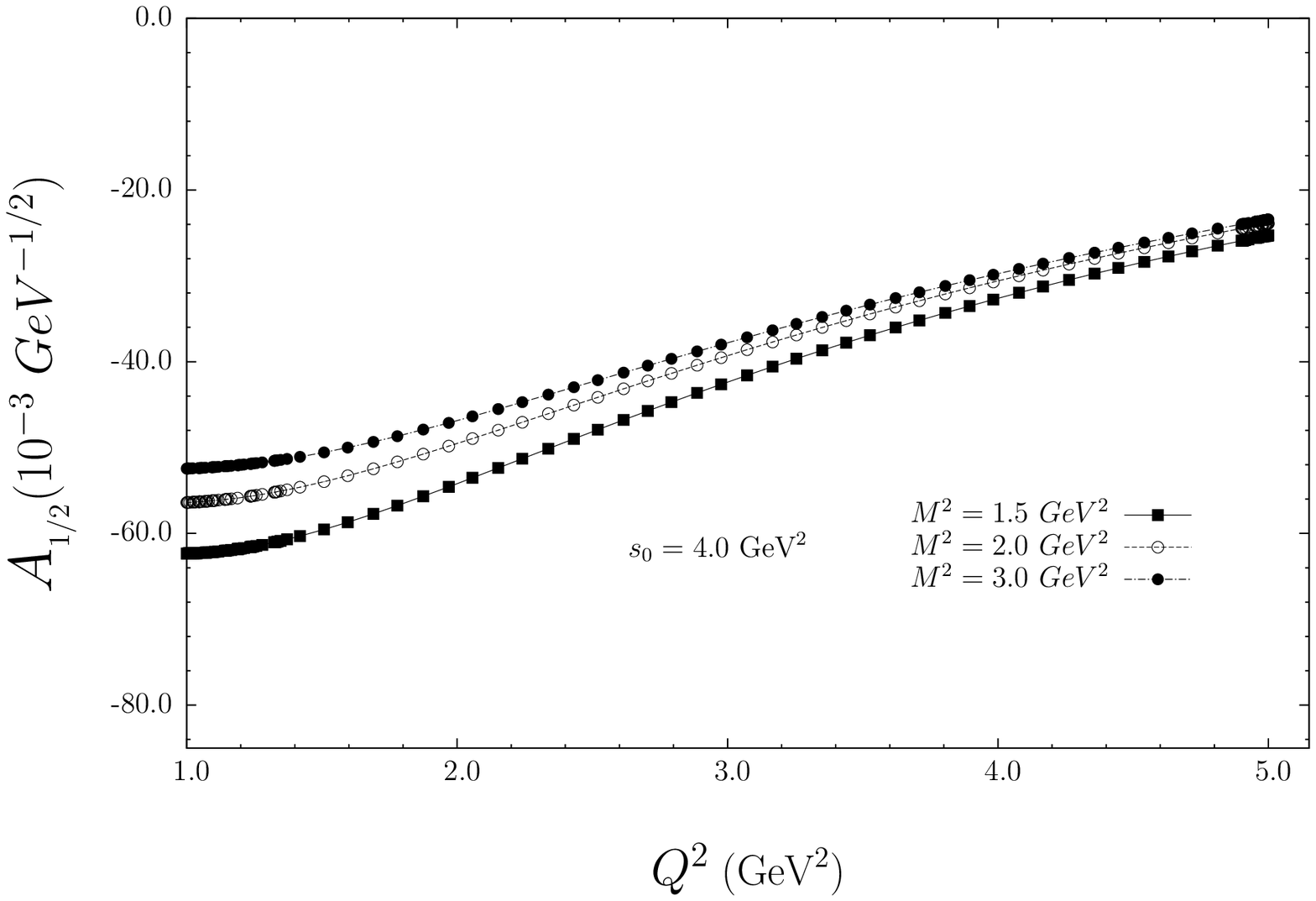}}
\end{center}
\vskip -12.0cm
\caption{}
\end{figure}

\begin{figure}[b]   
\vskip -1.5cm
\begin{center}
\scalebox{0.785}{\includegraphics{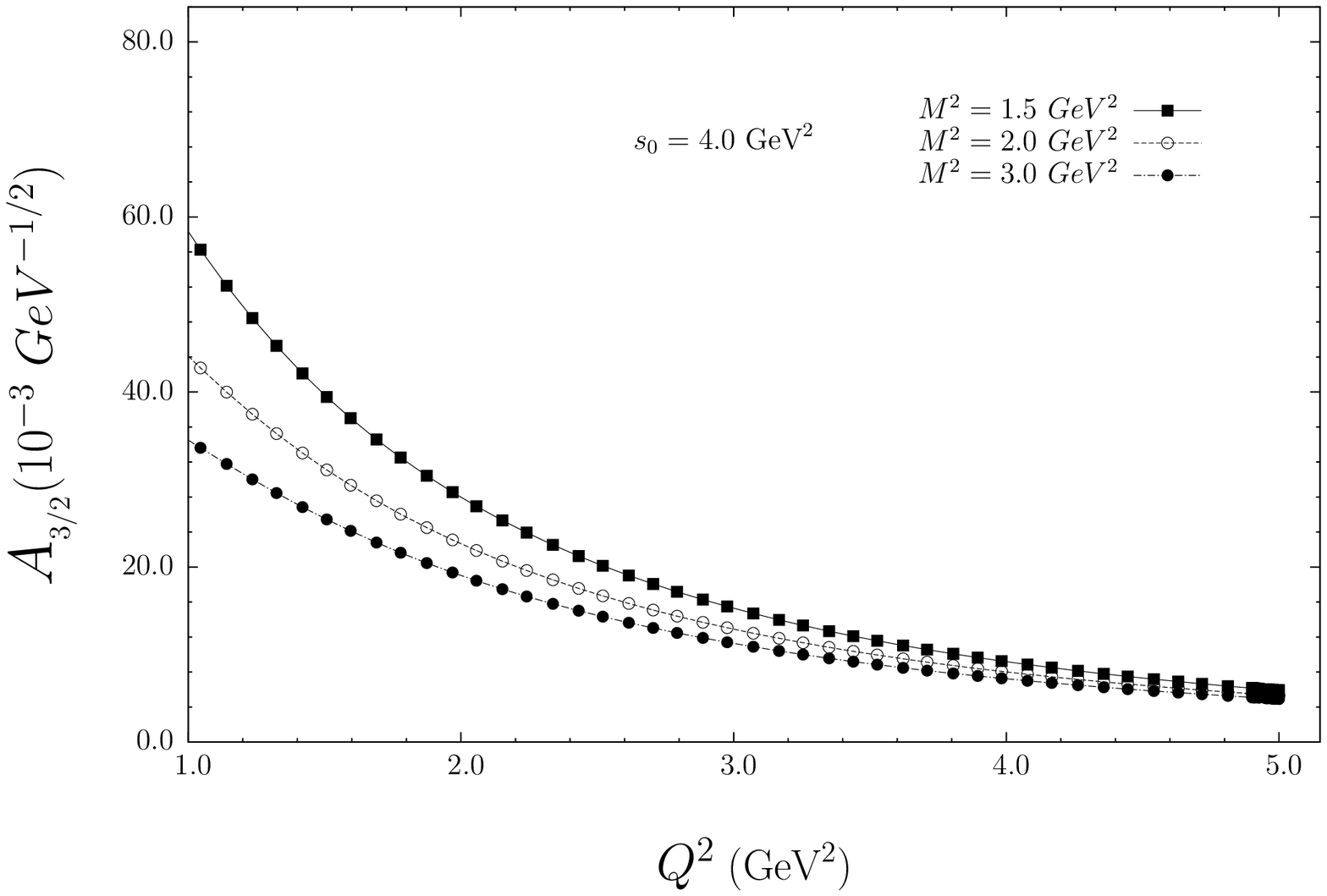}}
\end{center}
\vskip -12.0cm
\caption{}
\end{figure}

\begin{figure}[t]
\vskip -1.5cm
\begin{center}
\scalebox{0.785}{\includegraphics{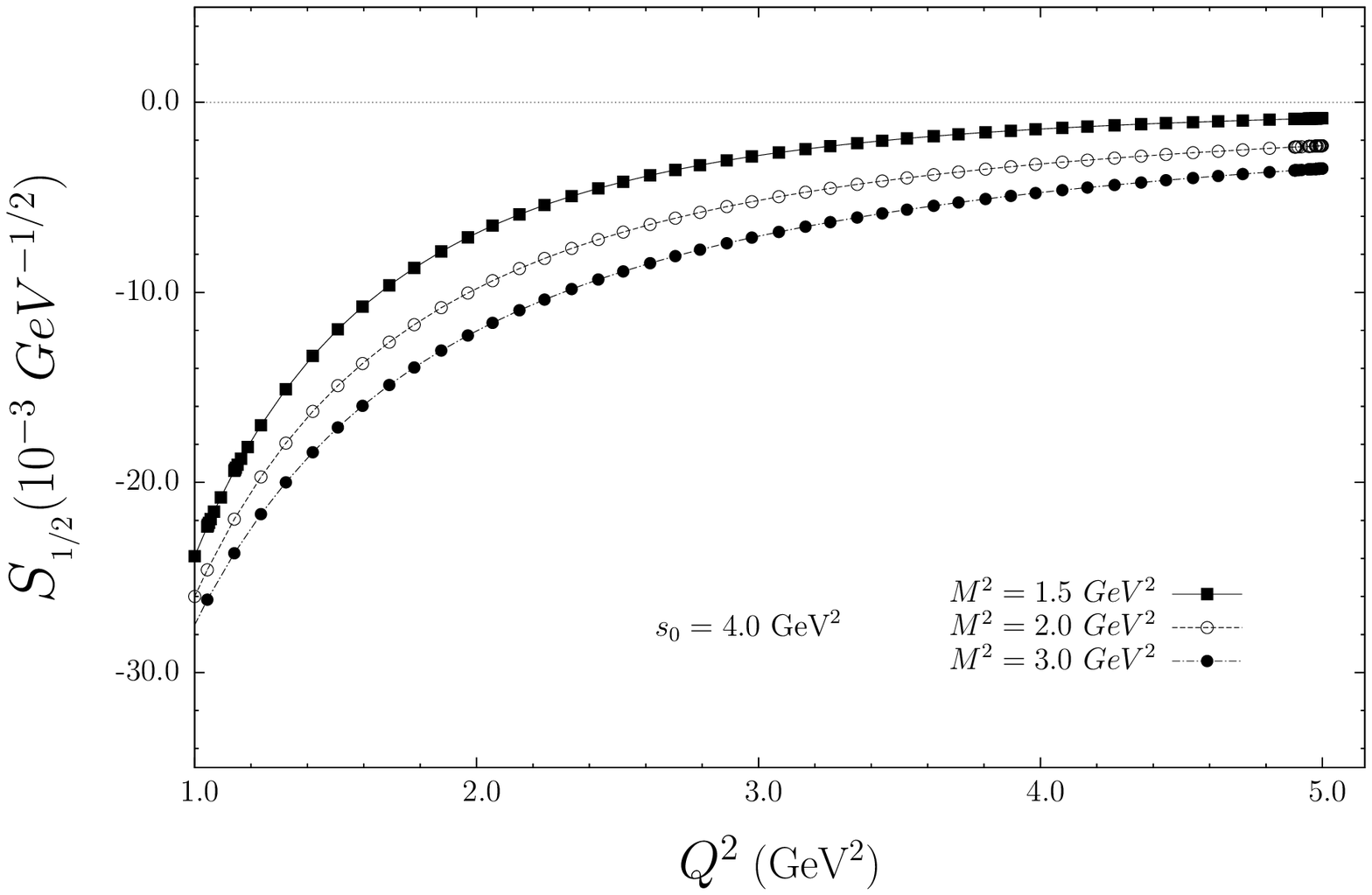}}
\end{center}
\vskip -12.0cm
\caption{}
\end{figure}

\end{document}